\documentclass[aps,prb,twocolumn,floatfix,superscriptaddress,showpacs,Superscript citations]{revtex4-1}

\usepackage{appendix}
\usepackage{graphicx}
\usepackage{amsfonts}
\usepackage{amsmath}
\usepackage{amssymb}
\usepackage{bm}
\usepackage{color}
\usepackage[hypertex]{hyperref}

\begin{document}

  \title{Nonuniform superconductivity and Josephson effect in conical ferromagnet}
  \author{Hao Meng}
  \affiliation{University Bordeaux, LOMA UMR-CNRS 5798, F-33405 Talence Cedex, France}
  \affiliation{School of Physics and Telecommunication Engineering, Shaanxi University of Technology, Hanzhong 723001, China}
  \affiliation{Shanghai Key Laboratory of High Temperature Superconductors, Shanghai University, Shanghai 200444, China}
  \author{A. V. Samokhvalov}
  \affiliation{Institute for Physics of Microstructures, Russian Academy of Sciences, 603950 Nizhny Novgorod, GSP-105, Russia}
  \affiliation{Lobachevsky State University of Nizhny Novgorod, Nizhny Novgorod 603950, Russia}
  \author{A. I. Buzdin}
  \email{alexandre.bouzdine@u-bordeaux.fr}
  \affiliation{University Bordeaux, LOMA UMR-CNRS 5798, F-33405 Talence Cedex, France}
  \affiliation{Department of Materials Science and Metallurgy, University of Cambridge, CB3 0FS, Cambridge, United Kingdom}
  \affiliation{Sechenov First Moscow State Medical University, Moscow, 119991, Russia}


  \begin{abstract}
   Using the Gorkov equations, we provide an exact solution for a one-dimensional model of superconductivity in the presence of a conical helicoidal exchange field. Due to the special type of symmetry of the system, the superconducting transition always occurs into a nonuniform superconducting phase (in contrast with the Fulde-Ferrell-Larkin-Ovchinnikov state, which appears only at low temperatures). We directly demonstrate that the uniform superconducting state in our model carries a current and thus does not correspond to the ground state. We study in the framework of the Bogoliubov-de Gennes approach the properties of the Josephson junction with a conical ferromagnet as a weak link. In our numerical calculations, we do not use any approximations (such as, e.g., a quasiclassical approach), and we show a realization of an anomalous $\phi_{0}$ junction (with a spontaneous phase difference $\phi_{0}$ in the ground state). The spontaneous phase difference $\phi_{0}$ strongly increases at high values of the exchange field near the borderline with a half-metal, and it exists also in the half-metal regime.
  \end{abstract}

   \maketitle

   \section{Introduction}

    The interest in superconductor-ferromagnet (SF) structures has been stimulated by the unusual SF proximity effect, leading to the fabrication of the Josephson junctions with unique properties (see, e.g.,~\cite{AAGolubov,Buz,BerRMP,LinderRobinson,MEschrig}), which paved the way for superconducting spintronics. Moreover, the combination of spin-orbit coupling and a Zeeman field may lead to the anomalous Josephson effect---the so-called $\phi_{0}$ junction with a spontaneous phase difference at the ground state~\cite{IVKrive1,AAReynoso,AIBuzdin,SMAB}. This is related to an emergence of topological nonuniform superconducting phases~\cite{Konstantin}. In \cite{Martin} it has been noted that a superconductor with a conical helical magnet structure is described by the same Hamiltonian as a topological superconducting phase appearing in systems with spin-orbit and Zeeman interactions.

    The problem of a superconducting uniform phase in the presence of the helicoidal exchange field has a complete analytical solution in the framework of the formalism of Gorkov's Green functions~\cite{LNBulaevskii2}. In \cite{MiodragLKulic} the peculiar properties of the Josephson junction between two helicoidal superconductors were considered, while in \cite{LevBulaevskii,AFVolkov,IVBob,DSRab,DSRabIVBob} the Josephson junction with a magnetic helix weak link was studied in the framework of the quasiclassical approximation.

    In Sec. II of this paper, we use Gorkov's formalism to get the analytical expressions for Green's functions in the conical helical superconducting magnet, taking into account the possibility of the topological nonuniform superconducting phase realization. Further, we perform a detailed analysis of the one-dimensional (1D) system and demonstrate the emergence of the nonuniform superconducting phase with a modulation wave vector $q$ when the helix becomes conical. The modulation vector is proportional to the canting of the helix and inversely proportional to the helix period. Our conclusion is based on the analysis of the critical temperature dependence on the superconductivity modulation vector $q$, which is obtained from the linear equation for the superconducting order parameter. The modulated superconducting state corresponds to the minimum energy of the system and does not carry current. Complimentarily, we calculate the current at $T=0$ in the uniform superconducting phase and show that it is not equal to zero, which proves that the uniform phase cannot be a ground state and thus the modulated phase is the most stable at all temperatures.

    The emergence of the modulated superconducting state may be illustrated by simple arguments in the framework of Ginzburg-Landau theory. In the standard situation, the lowest over the gradients of the order parameter $\Psi$ term gives the following well known quadratic contribution to the free energy, $\delta{F_{in\hom}}=\gamma\left\vert\mathbf{\nabla}\Psi\right\vert^{2}$, while the higher derivative terms may be neglected. The term that is linear over the gradient is absent because it is not invariant under the inversion symmetry operation. In the absence of inversion symmetry, Rashba spin-orbit interaction (SO) leads to the following additional contribution to the electron's energy: $\sim[\vec{\sigma}\times\vec{p}]\cdot\vec{n}$, where $\vec{p}$ is the momentum, $\vec{n}$ is the unit vector along the axis with broken inversion symmetry, and $\vec{\sigma}=\left(\sigma_{x},\sigma_{y},\sigma_{z}\right) $ is the vector of Pauli matrices~\cite{Mineev_Review}. In the presence of the exchange field $\vec{h}$ this results in a term that is linear over the gradient of the superconducting order parameter $\Psi$ in the Ginzburg-Landau (GL) free energy $\sim[\vec{n}\times \vec{h}]\cdot(\nabla\Psi)\,\Psi^*$ (see, for example, \cite{Mineev_Review,Edelstein-JPCM96}). In the case of the conical helicoid, the role of the $\vec{n}$ vector is played by the vector $[\,\vec{h} \times \mathrm{rot}\vec{h}\,]$, and the linear-over-gradient term becomes $\sim\left[\,\vec{h} \times [\,\vec{h}\times \mathrm{rot}\vec{h}]\,\right]\,\cdot(\nabla\Psi)\,\Psi^*$. This is a manifestation of the equivalence of a model of a conical superconductor to a model of a topological superconductor~\cite{Martin}. In the considered case of the conical helicoid with the exchange field $\vec{h}=(h\cos \mathbf{Q r},h\sin \mathbf{Q r},h_{z})$ (the wave vector $\mathbf{Q}=Q \mathbf{z}_0$ is along the $z$ axis), the normal state is lacking inversion symmetry and the following additional invariant that is linear over the gradient is possible:
    \begin{equation}\label{AddTerm}
       \delta F_{add}=
       \left[ i\lambda \Psi \left( \mathbf{h}\cdot rot\mathbf{h}\right)
       \cdot (h_{z}\cdot \left( \mathbf{\nabla }\Psi \right)_{z})+c.c.\right]\,,
    \end{equation}
    where the parameter $\lambda$ depends on the strength of the SO coupling. In the result, the energy contribution due to the modulation of the order parameter $\Psi=\Psi_{0}e^{iqz}$ becomes $\delta{F_{in\hom}}=\gamma{q^{2}}\Psi_{0}^{2}-2\lambda{q}h^{2}h_{z}Q\Psi_{0}^{2}$, and the minimum energy (and the maximum of the critical temperature) corresponds to the nonuniform superconducting state with a modulation vector $q\sim{h^{2}}h_{z}Q$. Note that there is no threshold on the value of counting field $h_{z}$ to generate the modulation, which is in sharp contrast with a Fulde-Ferrell-Larkin-Ovchinnikov (FFLO) state~\cite{FuldeFerrell,LarkinOvchinnikov}. The FFLO modulated state appears when the usual gradient term in the Ginzburg-Landau functional changes its sign, i.e., the coefficient $\gamma$ becomes negative when the exchange field overcomes some threshold~\cite{BuzdinKachkachi}.

     At low temperature for a standard superconductor we may use the London theory, and the gauge invariance imposes the following form of the term in the energy, depending on the vector-potential $\mathbf{A}$:
     $$
     \triangle{F}=a\left(\nabla\varphi+\frac{2e}{\hbar}\mathbf{A}\right)^2\,,
     $$
    where $\varphi$ is the phase of the superconducting order parameter $\Psi=|\Psi|\,\exp{(i\varphi})$. As a consequence, the current density $j=-c\, \delta F/\delta\mathbf{A}$, and in the absence of the magnetic field, choosing $A =0$, we see that the minimum energy corresponds to $\nabla\varphi=0$ and therefore $j=0$. In the considered case of the conical helicoid, the contribution $\Delta{F}$ to the energy should have a linear over $\left(\nabla\varphi+2e\mathbf{A}/\hbar\right)$ term:
    \begin{equation}\label{LondonEner}
       \triangle{F}=a\left(\nabla\varphi+\frac{2e}{\hbar}
       \mathbf{A}\right)^2+b\left(\nabla\varphi+\frac{2e}{\hbar}
       \mathbf{A} \right)\,.
    \end{equation}

    As a result, the current density reads
    $$
    j\sim2a\left(\nabla\varphi+\frac{2e}{\hbar}
    \mathbf{A}\right)+b\,,
    $$
    and in the absence of the field ($A=0$) and phase modulation ($\nabla\varphi=0$) the current is nonzero, $j\sim{b}$. This reflects the fact that the uniform state is not a ground-state of our system. Indeed, for $A=0$ the minimum of the energy (\ref{LondonEner}) corresponds to $\nabla\varphi=-b/2a$, and for this phase modulation the current vanishes.

    In Sec. III we calculate the Josephson current for the 1D model of the weak link made of the conical helix. Our numerical calculations use the exact solutions of the Bogoliubov-de Gennes (BdG) equations and we demonstrate the realization of the anomalous $\phi_{0}$ junction. The spontaneous phase shift $\phi_{0}$ strongly increases when we approach the half-metal regime or when we are completely in the half-metal state. In this case, the current-phase relation for the supercurrent is $I(\phi)=I_{c}\sin(\phi-\phi_{0})$ and the additional phase shift $\phi_{0}$ is proportional to the ferromagnetic component of exchange field $h_{z}$. We provide a detailed study of the properties of the $\phi_{0}$ junction as a function of conical helix parameters. The conical helicoidal phase exists, for example, in antiferromagnetic Ho, and the Ho/Nb structure has attracted a lot of attention~\cite{BHalasz,ISosnin,JDSWitt,FChiodi,ADBernardo}. In these systems, the electron mean free path is of the same order as the period of the helix, and we believe that qualitatively the results of our work may be applicable to these structures. The possibility to use the conical helix as a building block of the $\phi_{0}$ junction may be important for the design of the superconducting spintronics devices.

    \begin{figure}
    \centering
   \includegraphics[width=3.3in]{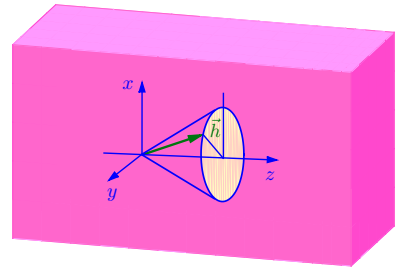} 
    \caption{The sketch of a superconductor with a conical magnetic texture. The green thick arrow indicates the direction of the exchange field.}\label{fig1}
   \end{figure}

   \section{Superconducting conical helicoidal phase---Gorkov's Green Functions}

    We study a clean \emph{s}-wave magnetic superconductor with conical magnetic order. The conical magnetism and the spatially modulated order parameter can be characterized by $\vec{h}=(h\cos\mathbf{Q\cdot{r}},h\sin\mathbf{Q\cdot{r}},h_{z})$ and $\Delta(\mathbf{r})=\Delta{e^{i\mathbf{q\cdot{r}}}}$, respectively (see Fig.~\ref{fig1}). Using the mean-field approximation, we may write the Hamiltonian of the system as~\cite{PGdeGennes}
    \begin{align}
        \hat{H}&=\sum_{\alpha\beta}\int{d^{3}}\mathbf{r}\left\{\hat{\psi}_{\alpha }^{\dagger}(\mathbf{r})\xi_{p}\hat{\psi}_{\alpha}(\mathbf{r})+
        \hat{\psi}_{\alpha}^{\dagger}(\mathbf{r})(\vec{h}\cdot\vec{\sigma})_{\alpha\beta}\hat{\psi}_{\beta}(\mathbf{r})\right.  \label{Hamilt_BCS} \\
        &\left.+\frac{1}{2}\left[(i\sigma_{y})_{\alpha\beta}\Delta(\mathbf{r})\hat{\psi}_{\alpha}^{\dagger}(\mathbf{r}) \hat{\psi}_{\beta}^{\dagger}(\mathbf{r})+h.c\right]\right\}, \notag
    \end{align}
    where $\xi_{p}=\frac{p^{2}}{2m}-E_{F}$, and $\hat{\psi}_{\alpha}^{\dagger}(\mathbf{r})$ and $\hat{\psi}_{\alpha}(\mathbf{r})$ represent creation and annihilation operators with spin $\alpha$. The spatially modulated superconducting order parameter is described by $\left\langle\hat{\psi}_{\alpha}^{\dagger}(\mathbf{r})\hat{\psi}_{\beta}^{\dagger}(\mathbf{r})\right\rangle=\left(i\sigma_{y}\right)_{\alpha\beta}\Delta{e^{i\mathbf{q\cdot{r}}}}$. The Gorkov equations of the system of the Green's functions $G_{\alpha,\beta}(\mathbf{r},\mathbf{r}^{\prime})=-\left\langle{T}\hat{\psi}_{\alpha}(\mathbf{r})\hat{\psi}_{\beta}^{\dagger}(\mathbf{r}^{\prime})\right\rangle$ and $F_{\alpha,\beta}^{\dagger}(\mathbf{r},\mathbf{r}^{\prime})=\left\langle{T}\hat{\psi}_{\alpha}^{\dagger}(\mathbf{r})\hat{\psi}_{\beta}^{\dagger}(\mathbf{r}^{\prime})\right\rangle$ have the form
    \begin{equation}
       \left(i\omega_{n}-\xi_{p}-\hat{V}\right)\hat{G}(\mathbf{r},\mathbf{r}^{\prime})+\Delta{e^{i\mathbf{q\cdot{r}}}}\cdot {\hat{I}}\hat{F}^{\dag}(\mathbf{r},\mathbf{r}^{\prime})=\delta(\mathbf{r}-\mathbf{r}^{\prime}),  \label{Gorkov1}
    \end{equation}
    \begin{equation}
       \left(i\omega_{n}+\xi_{p}+\tilde{V}\right)\hat{F}^{\dag}(\mathbf{r},\mathbf{r}^{\prime})-\Delta^{\ast}{e^{-i\mathbf{q\cdot{r}}}}\cdot {\hat{I}}\hat{G}(\mathbf{r},\mathbf{r}^{\prime})=0,  \label{Gorkov2}
    \end{equation}
    where the matrix $\hat{I}$ is written as
    \begin{equation}
       \hat{I}=i\sigma_{y}=\left(
       \begin{array}{cc}
           0 & 1 \\
          -1 & 0
       \end{array}
       \right).  \label{matrix_I}
    \end{equation}
    The wave vectors $\mathbf{Q}$ and $\mathbf{q}$ are along the $z$-axis, and the potential of the conical magnetic order is given by
    \begin{equation}
         \hat{V}(\mathbf{r})=\vec{h}\cdot\vec{\sigma}=\left(
         \begin{array}{cc}
             h_{z} & he^{-iQz} \\
             he^{iQz} & -h_{z}
         \end{array}
         \right)    \label{matrix_V1}
     \end{equation}
    while
    \begin{equation}
        \tilde{V}(\mathbf{r})=\left(
        \begin{array}{cc}
            h_{z} & he^{iQz} \\
            he^{-iQz} & -h_{z}
        \end{array}
        \right). \label{matrix_V2}
    \end{equation}

    Using the Fourier transform, we obtain the exact solution of (\ref{Gorkov1})-(\ref{Gorkov2}) described in Appendix~\ref{AppendA} and get the Green functions (below only $\hat{F}_{21}^{\dag}$ and $\hat{G}_{11}$ are presented)
    \begin{gather}
        \hat{F}_{21}^{\dag}\left(p-\frac{Q}{2}-\frac{q}{2},p^{\prime}\right)
        =-\delta\left(p-\frac{Q}{2}+\frac{q}{2}-p^{\prime}\right)   \label{F21}  \\
        \times\frac{\left[\left(i\omega_{n}-\xi_{2}+h_{z}\right)\left(i\omega_{n}+\xi_{3}+h_{z}\right)+h^{2}-\left\vert\Delta\right\vert ^{2}\right]\Delta^{^{\ast}}}{D(\omega_{n})},  \notag
    \end{gather}
    \begin{gather}
        \hat{G}_{11}\left(p-\frac{Q}{2}+\frac{q}{2},p^{\prime}\right)=\delta\left(p-\frac{Q}{2}+\frac{q}{2}-p^{\prime}\right)  \label{G11} \\
        \times\left[\frac{\left(i\omega_{n}-\xi_{2}+h_{z}\right)\left(i\omega_{n}+\xi_{3}+h_{z}\right)\left(i\omega_{n}+\xi_{4}-h_{z}\right)}{D(\omega_{n})}\right.  \notag \\
        \left.-\frac{\left(i\omega_{n}-\xi_{2}+h_{z}\right)h^{2}+\left(i\omega_{n}+\xi_{4}-h_{z}\right)\left\vert\Delta\right\vert^{2}}{D(\omega_{n})}\right], \notag
    \end{gather}
    where $D(\omega_{n})=$
    \begin{eqnarray}
       &&\left[\left(i\omega_{n}-\xi_{1}-h_{z}\right)\left(i\omega_{n}+\xi_{4}-h_{z}\right)+h^{2}-\left\vert\Delta\right\vert^{2}\right] \label{DW} \\
       &&\times\left[\left(i\omega_{n}-\xi_{2}+h_{z}\right)\left(i\omega_{n}+\xi_{3}+h_{z}\right)+h^{2}-\left\vert\Delta\right\vert^{2}\right] \notag \\
       &&-\left(2i\omega_{n}-\xi_{1}+\xi_{3}\right)\left(2i\omega_{n}-\xi_{2}+\xi_{4}\right)h^{2}.  \notag
    \end{eqnarray}
     and
    \begin{equation}
        \xi_{1}=\xi_{p-\frac{Q}{2}+\frac{q}{2}}, \xi _{2}=\xi_{p+\frac{Q}{2}+\frac{q}{2}}, \\
    \end{equation}
    \begin{equation}
        \xi_{3}=\xi_{p+\frac{Q}{2}-\frac{q}{2}}, \xi_{4}=\xi_{p-\frac{Q}{2}-\frac{q}{2}}.
    \end{equation}

    Note that we have obtained the exact solution of the 1D model, which is readily generalized to the 3D case: indeed we start from the Hamiltonian (\ref{Hamilt_BCS}) describing the 3D system, and the corresponding Gorkov equations (\ref{Gorkov1}) and (\ref{Gorkov2}) are readily applied to the 3D case provided that we consider all vectors as 3D vectors $\vec{p}$, $\vec{Q}$, and $\vec{q}$. The superconducting conical ferromagnet is one of the rare examples when it is possible to get explicitly the complete solution in the framework of the microscopical Gorkov equations.

    \subsection{The energy spectrum of the conical ferromagnet}

\begin{figure}
     \centering
     \includegraphics[width=3.3in]{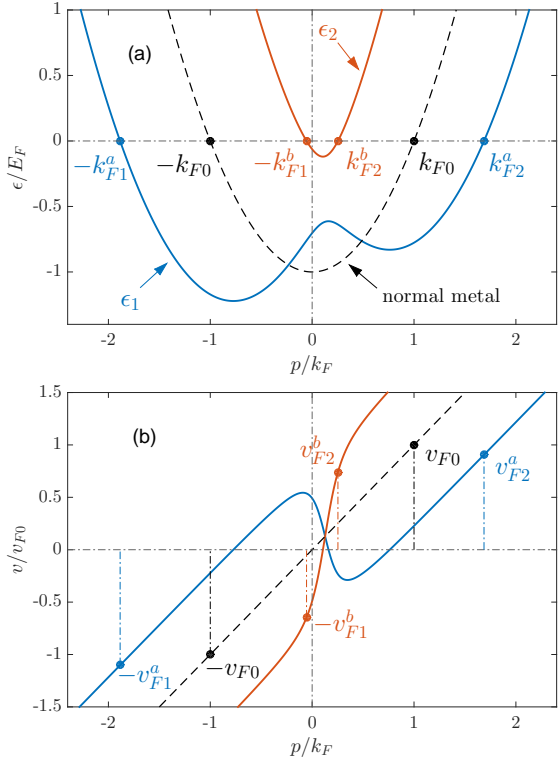} 
      \caption{(a) Energy spectrums ($\protect\epsilon_{1}$ and $\protect\epsilon_{2}$) for the conical ferromagnet ($h/E_{F}=0.25$, $h_{z}/E_{F}=0.2$, and $Q/k_{F}=\protect\pi/2$) and the normal metal ($h/E_{F}=0$, $h_{z}/E_{F}=0$, and $Q/k_{F}=0$); (b) the velocity of a quasiparticle as a function of the wave vector $p$. Here the quantities of the velocity are normalized to the value of the Fermi velocity $v_{F0}$ of the quasiparticle in the normal metal.}     \label{fig2}
     \end{figure}

    Let us first consider the normal conical ferromagnet without superconducting coupling ($\Delta=0$ and $q=0$). The Green's function $\hat{G}_{11}$ in such a case reads
    \begin{gather}
       \hat{G}_{11}\left(p-\frac{Q}{2},p^{\prime}\right)=\delta\left(p-\frac{Q}{2}-p^{\prime}\right)  \label{G11f}  \\
       \times\frac{i\omega_{n}-\xi_{p+\frac{Q}{2}}+h_{z}}{\left(i\omega_{n}-\xi_{p-\frac{Q}{2}}-h_{z}\right)\left(i\omega_{n}-\xi_{p+\frac{Q}{2}}+h_{z}\right)-h^{2}}.  \notag
    \end{gather}
    To find the electrons spectrum $\epsilon$, we should perform the analytical continuation $i\omega_{n}\rightarrow\epsilon$ in the denominator of the equation (\ref{G11f}), and then its zeros give us the equation for the energy spectrum,
    \begin{equation}
       \left(\epsilon-\xi_{p-\frac{Q}{2}}-h_{z}\right)\left(\epsilon-\xi_{p+\frac{Q}{2}}+h_{z}\right)-h^{2}=0. \label{spectrum_eq}
    \end{equation}
    In the result, we obtain two branches of the energy spectrum,
    \begin{eqnarray}
       \epsilon_{1(2)}&=&\frac{1}{2}\left[\xi _{p-\frac{Q}{2}}+\xi_{p+\frac{Q}{2}}\right.  \label{Ener_spec}  \\
       &&\mp\left.\sqrt{\left(\xi_{p+\frac{Q}{2}}-\xi_{p-\frac{Q}{2}}-2h_{z}\right)^{2}+4h^{2}}\right].    \notag
    \end{eqnarray}
     As illustrated in Fig.~\ref{fig2}(a), two branches ($\epsilon_{1}$ and $\epsilon_{2}$) of the energy spectrum are not symmetric with respect to $p=0$, and they also do not contain the gaps. It is a peculiar property of the periodical helicoidal exchange field---it does not create the gap band structure in contrast to the usual case of the periodical potential field.

     According to the formula $v=\frac{1}{\hbar}\frac{dE}{dk}$, we can compute the velocity of quasiparticles [see Fig.~\ref{fig2}(b)]. It is known that in the normal metal, the Fermi velocities of two quasiparticles (at $\pm{k_{F0}})$ have the same absolute values $v_{F0}$ of the Fermi velocities. However, in the conical ferromagnet, the absolute values of Fermi velocities of the quasiparticles are different in the same branches, for instance in the $\epsilon_{1}$ branch ($v_{F1}^{a}=1.102v_{F0}$ and $v_{F2}^{a}=0.9057v_{F0}$) and in the $\epsilon_{2}$ branch ($v_{F1}^{b}=0.6467v_{F0}$ and $v_{F2}^{b}=0.7397v_{F0}$) for chosen parameters of the conical ferromagnet. Namely, this property is characteristic of the systems with a spin-orbit interaction and leads to the appearance of the modulated superconducting states.

     \subsection{Superconducting transition temperature in the modulated phase}

    The critical temperature of the system is determined by the linearized self-consistency equation (taking in the limit $\Delta\rightarrow{0}$):
    \begin{equation}
        \Delta^{\ast}=|g|T\sum_{\omega_{n}}\int_{-\infty}^{+\infty}\hat{F}_{21}^{\dag}\frac{dp}{2\pi}, \label{self-consistency1}
    \end{equation}
    where $g$ is the electron-phonon coupling constant. It is more convenient to write it in the following form:
    \begin{equation}
       \ln\left(\frac{T_{c}}{T_{c0}}\right)=2T_{c}\sum_{\omega_{n}\geq0}\left[{\rm Re}\int_{-\infty}^{+\infty} \frac{\hat{F}_{21}^{\dag}}{\Delta^{\ast}}d\xi-\frac{\pi}{\omega_{n}}\right], \label{self-consistency2}
    \end{equation}
    where $T_{c}$ is the critical temperature and $T_{c0}$ is the critical temperature in the absence of exchange field $\vec{h}$. Introducing $\Delta{T_{c}}=T_{c}-T_{c0}$, and performing the expansion over the modulation vector $q$ of the superconducting phase in the limit $h_{z}$, $h\ll{T_{c}}$ we finally obtain (see Appendix \ref{AppendB} for details)
    \begin{eqnarray}
        \frac{\Delta{T_{c}}}{T_{c0}}&=&2\pi{T_{c}}\sum_{\omega_{n}\geq0}\left[-\frac{h_{z}^{2}}{\omega_{n}^{3}} -\frac{4h^{2}}{\left(4\omega_{n}^{2}+v^{2}Q^{2}\right)\omega_{n}}\right.   \label{Tc}  \\
        &&\left.-\frac{4Qh^{2}h_{z}q}{m\left(4\omega_{n}^{2}+v^{2}Q^{2}\right)\omega_{n}^{3}}-\frac{v^{2}q^{2}}{4\omega_{n}^{3}}\right].  \notag
    \end{eqnarray}

    The very important point is the presence of linear-over-$q$ term, which means that the maximum of the critical temperature always occurs at finite $q$. The linear dependence of the critical temperature $T_c$ over $q$ (which describes the modulation of the superconducting order parameter) is the direct consequence of the linear-over-gradient term $\nabla\Psi$ in the GL free energy (\ref{AddTerm}). In accordance with the form of the GL term, the coefficient on $q$ dependence is proportional to the product $h^2h_zQ$. At the same time, the presence of a linear-over-$q$ term guarantees that the modulated state corresponds to the absence of the current, while the uniform one ($q=0$) does not.

    For $vQ\ll{T_{c0}}$, the above equation can be simplified as
    \begin{equation}
       \frac{\Delta{T_{c}}}{T_{c0}}=2\pi{T_{c}}\sum_{\omega_{n}\geq0}\left[-
       \frac{h_{z}^{2}+h^{2}}{\omega_{n}^{3}}-\frac{Qh^{2}h_{z}}{m\omega_{n}^{5}}q-\frac{v^{2}}{4\omega_{n}^{3}}q^{2}\right] .
    \end{equation}
    The maximum of the transition temperature is reached at the modulation wave vector
    \begin{equation}
        q_{0}=-\frac{31Qh^{2}h_{z}}{28\pi^{2}T_{c0}^{2}E_{F}}\frac{\zeta(5)}{\zeta(3)}.  \label{max_q}
    \end{equation}
    Here $\zeta(s)$ is the Euler--Riemann zeta function and $E_{F}=\frac{mv^{2}}{2}$.

    In the opposite limit, $vQ\gg T_{c0}$, the above equation will change to
    \begin{eqnarray}
        \frac{\Delta{T_{c}}}{T_{c0}}&=&2\pi{T_{c}}\sum_{\omega_{n}\geq0} \left[-\left(\frac{h_{z}^{2}}{\omega_{n}^{2}}+\frac{4h^{2}}{v^{2}Q^{2}}\right)\frac{1}{\omega_{n}}\right. \\
        &&\left.-\frac{4h^{2}h_{z}}{mv^{2}Q\omega_{n}^{3}}q-\frac{v^{2}}{4\omega_{n}^{3}}q^{2}\right]  \notag
    \end{eqnarray}
    and the modulation vector of the superconducting phase will be $q_{0}=-\frac{4h^{2}h_{z}}{v^{2}QE_{F}}$. Note that in the both cases, the expression for the modulation vector $q_{0}$ contains a small factor $\frac{h_{z}}{E_{F}}$, and this circumstance explains why the emergence of the modulated superconducting phase cannot be described in the framework of Eilenberger or Usadel quasiclassical equations, where such effects are simply neglected.

   \subsection{Current in a uniform superconducting phase with the conical magnetic order}

   \begin{figure}
   \centering
   \includegraphics[width=3.3in]{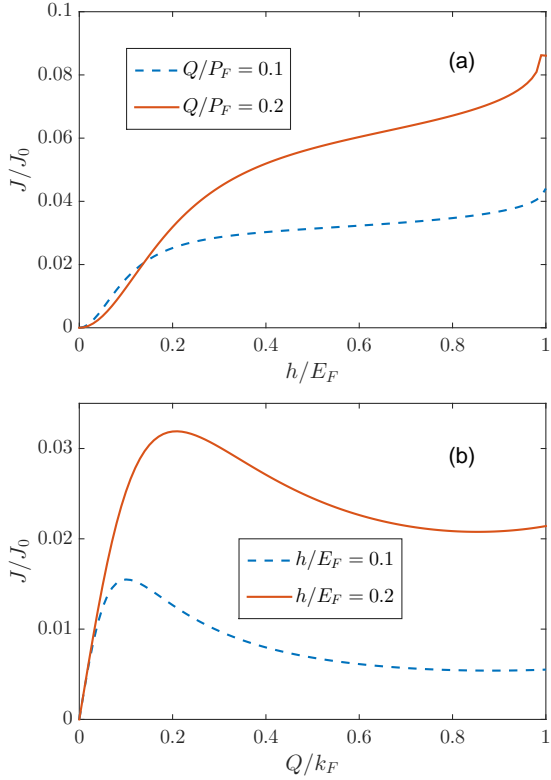} 
    \caption{The supercurrent $J$ versus (a) the magnetic order $h$ and (b) the spiral wave vector $Q$. We choose $E_{F}=100\Delta$ and $h_{z}/E_{F}=0.05$. Here the supercurrent unit is $J_{0}=\frac{2e}{m}$.}     \label{fig3}
    \end{figure}

     We now derive the expression for supercurrent in uniform ($q=0$) superconductors with the conical spiral magnetic order. The spiral magnetic order is characterized by the wave vector $\mathbf{Q}$ along the $z$ axis, $\mathbf{Q}=\chi Q\mathbf{e}_{z}$, and by the helicity $\chi=\pm1$. In the limit $h_{z}\ll\left\vert\Delta\right\vert$, the Green function $\hat{G}_{11}\left(p-\frac{Q}{2},p^{\prime}\right)$ reads
     \begin{equation}
         \hat{G}_{11}\left(p-\frac{Q}{2},p^{\prime}\right)=\hat{G}_{11}^{(0)}\left(p-\frac{Q}{2},p^{\prime}\right) +h_{z}\hat{G}_{11}^{(1)}\left(p-\frac{Q}{2},p^{\prime}\right),    \label{G11P}
     \end{equation}
      where
     \begin{eqnarray}
         &&\hat{G}_{11}^{(0)}\left(p-\frac{Q}{2},p^{\prime}\right)=\delta\left(p-\frac{Q}{2}-p^{\prime}\right)  \\
         &&\times\frac{-\xi_{p-\frac{Q}{2}}\left(\omega_{n}^{2}+\xi_{p+\frac{Q}{2}}^{2}+\left\vert\Delta\right\vert^{2}\right) -\xi_{p+\frac{Q}{2}}h^{2}}{\left[\omega_{n}^{2}+E_{1}^{2}\right]\left[\omega_{n}^{2}+E_{2}^{2}\right]},  \notag
     \end{eqnarray}
     \begin{eqnarray}
         &&\hat{G}_{11}^{(1)}\left(p-\frac{Q}{2},p^{\prime}\right)=\delta\left(p-\frac{Q}{2}-p^{\prime}\right) \\
         &&\times\left\{\frac{\xi_{p+\frac{Q}{2}}^{2}-\omega_{n}^{2}-h^{2}+\left\vert\Delta\right\vert^{2}}{\left[\omega_{n}^{2} +E_{1}^{2}\right]\left[\omega_{n}^{2}+E_{2}^{2}\right]}\right.   \notag \\
         &&\left.-2\omega_{n}^{2}\frac{\left(\xi_{p+\frac{Q}{2}}^{2}-\xi_{p-\frac{Q}{2}}^{2}\right)\left(\omega_{n}^{2} +\xi_{p+\frac{Q}{2}}^{2}+\left\vert\Delta\right\vert^{2}+h^{2}\right)}{\left[\omega_{n}^{2} +E_{1}^{2}\right]^{2}\left[\omega_{n}^{2}+E_{2}^{2}\right]^{2}}\right\}, \notag
     \end{eqnarray}
     \begin{eqnarray}
         E_{1,2}^{2}&=&\tilde{\zeta}^{2}+\tilde{\eta}^{2}+\left\vert\Delta\right\vert^{2}+h^{2} \\
         &&\pm2\sqrt{\tilde{\zeta}^{2}\left(\tilde{\eta}^{2}+h^{2}\right)+\left\vert\Delta\right\vert^{2}h^{2}},  \notag
     \end{eqnarray}
     $\tilde{\zeta}=\left(\xi_{p-\frac{Q}{2}}+\xi_{p+\frac{Q}{2}}\right)/2$ and $\tilde{\eta}=\left(\xi_{p-\frac{Q}{2}}-\xi_{p+\frac{Q}{2}}\right)/2$.

     \begin{figure}
        \centering
        \includegraphics[width=3.3in]{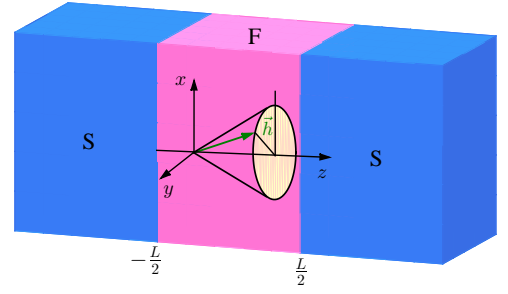} 
        \caption{The SFS Josephson junction consists of two \emph{s}-wave superconductors and a conical ferromagnet. The green thick arrow indicates the direction of the exchange field in the conical ferromagnet.}  \label{fig4}
     \end{figure}

     We may write for the current~\cite{Abrikosov}
     \begin{eqnarray}
         J&=&\frac{ie}{m}\left(\nabla_{r^{\prime}}-\nabla_{r}\right)\left.\left[\hat{G}_{11}(r,r^{\prime}) +\hat{G}_{22}(r,r^{\prime})\right]\right\vert_{r^{\prime}\rightarrow{r}}     \label{currentJ} \\
         &=&\frac{2e}{m}\iint{dp}d\omega\left[p\hat{G}_{11}(p,h_{z})+p\hat{G}_{22}(p,h_{z})\right]  \notag \\
         &=&\frac{4eh_{z}}{m}\left\{\frac{Q\pi}{m}\int{dp}\frac{p^{2}\xi_{Q}(p)}{E_{1}^{2}E_{2}+E_{1}E_{2}^{2}}\right.  \notag \\
         &&+\frac{Q\pi}{2}\int{dp}\frac{2h^{2}-\left(E_{1}-E_{2}\right)^{2}/2}{E_{1}^{2}E_{2}+E_{1}E_{2}^{2}}  \notag \\
         &&\left.+\frac{Q\pi}{m}\int{dp}\frac{p^{2}\xi_{Q}(p)\left[\frac{Q^{2}}{m}\xi_{Q}(p) -\left(E_{1}+E_{2}\right)^{2}\right]}{E_{1}E_{2}(E_{1}+E_{2})^{3}}\right\},  \notag
     \end{eqnarray}
     where $\xi_{Q}(p)=\xi(p)+\frac{Q^{2}}{8m}$. The details of these calculations are presented in Appendix \ref{AppendC}. From the above formula (\ref{currentJ}), we can obtain the dependence of supercurrent $J$ on the strength of the helical field $h/E_{F}$ and the helix wave vector $Q/k_{F}$ (see Fig.~\ref{fig3}). We see that the current in the uniform state is proportional to $h_{z}h^{2}$ and the spiral wave vector $Q$ in accordance with the results of Sec. IIB. Therefore, the uniform superconducting phase is not a ground state, which should be a nonuniform superconducting phase at any temperatures.

     \section{The Bogoliubov--de Gennes approach for conical Josephson junction}

     It is known that the effects related to the spin-orbit interaction often cannot be adequately described by the usual quasiclassical approach~\cite{ChristopherR,MASilaev}. As mentioned beforehand, the superconductor with a conical helical magnetic structure is similar to the topological superconducting phase appearing in the systems with spin-orbit and Zeeman interactions. So the anomalous supercurrent in the Josephson junction with conical magnetization should be calculated using exact solutions of the BdG approach but not the quasiclassical one.

     We consider the SFS Josephson junction made of two BCS superconductors (S) and a normal-state metal barrier (F) with conical magnetic spiral ordering, see Fig.~\ref{fig4}. The $z$ axis is chosen to be perpendicular to the layer interfaces with the origin located at the center of the ferromagnetic layer. The superconducting gap is supposed to be constant in the leads $\left(\left\vert{z}\right\vert>L/2\right)$ and absent inside the conical ferromagnet $\left(\left\vert{z}\right\vert<L/2\right)$:
     \begin{equation}
         \Delta (\mathbf{r})=\left\{
         \begin{array}{cc}
             \Delta \,e^{i\phi/2}\,, & z<-L/2\,, \\
             0\,, & |z|\leq{L/2} \,, \\
             \Delta \,e^{-i\phi/2}\,, & z>L/2,
         \end{array}
         \right.     \label{StepDelta}
     \end{equation}
     where $\Delta$ is the magnitude of the gap and $\phi$ is the phase difference between the two leads. As before, the spiral is characterized by the wave vector $\mathbf{Q}$ along the $z$ axis, $\mathbf{Q}=\chi{Q}\mathbf{e}_{z}$, and by the helicity $\chi=Q_{z}/Q=\pm{1}$. The BCS mean-field effective Hamiltonian of the considered system is described by the expression (\ref{Hamilt_BCS})~\cite{Buz,PGdeGennes} with a step-like $\Delta(\mathbf{z})$ (\ref{StepDelta}).

     To diagonalize the effective Hamiltonian, we use the Bogoliubov transformation $\hat{\psi}_{\alpha}(\mathbf{r})=\sum_{n}[u_{n\alpha}(\mathbf{r})\hat{\gamma}_{n}+v_{n\alpha}^{\ast}(\mathbf{r})\hat{\gamma}_{n}^{\dag}]$ and take into account the anticommutation relations of the quasiparticle annihilation operator $\hat{\gamma}_{n}$ and creation operator $\hat{\gamma}_{n}^{\dag}$. Using the presentation $u_{n\alpha}(\mathbf{r})=u_{p}^{\alpha}e^{ipz}$, $v_{n\alpha}(\mathbf{r})=v_{p}^{\alpha}e^{ipz}$, the resulting Bogoliubov-de Gennes (BdG) equations can be expressed as~\cite{PGdeGennes}
     \begin{equation}
         \begin{pmatrix}
             \hat{H}_{1} & i\hat{\sigma}_{y}\Delta (z) \\
             -i\hat{\sigma}_{y}\Delta^{\ast}(z) & -\hat{H}_{2}
         \end{pmatrix}
         \begin{pmatrix}
             \hat{u}(z) \\
             \hat{v}(z)
         \end{pmatrix}
         =\epsilon
         \begin{pmatrix}
             \hat{u}(z) \\
             \hat{v}(z)
         \end{pmatrix},   \label{BdG}
     \end{equation}
     where
     \begin{equation*}
         \hat{H}_{1(2)}=
         \begin{pmatrix}
              \xi_{p\mp{Q/2}}+{h_{z}} & {h} \\
              {h} & \xi_{p\pm{Q/2}}-h_{z}
         \end{pmatrix}.
     \end{equation*}
      Moreover, $\hat{u}(z)=[u_{p-Q/2}^{\uparrow}(z), u_{p+Q/2}^{\downarrow}(z)]^{T}$ and $\hat{v}(z)=[v_{p+Q/2}^{\uparrow}(z),v_{p-Q/2}^{\downarrow}(z)]^{T}$ are quasiparticle and quasihole wave functions, respectively.

      The solutions of the BdG equation (\ref{BdG}) can be found in each layer separately and then matched with the boundary conditions. For a given energy $\epsilon$ inside the superconducting gap, we find the following plane-wave solutions in the left superconducting electrode:
      \begin{eqnarray}
           \psi_{L}^{S}(z)&=&C_{1}\hat{\rho}_{1}e^{-ik_{S}^{+}z}+C_{2}\hat{\rho}_{2}e^{ik_{S}^{-}z}  \label{functionSL}  \\
           &&+C_{3}\hat{\rho}_{3}e^{-ik_{S}^{+}z}+C_{2}\hat{\rho}_{4}e^{ik_{S}^{-}z},   \notag
      \end{eqnarray}
      where $k_{S}^{\pm}=k_{F}\sqrt{1\pm{i}\sqrt{\Delta^{2}-\epsilon^{2}}/E_{F}}$ are the wave vectors for quasiparticles. $\hat{\rho}_{1}=[1,0,0,R_{1}e^{-i\phi/2}]^{T}$, $\hat{\rho}_{2}=[1,0,0,R_{2}e^{-i\phi/2}]^{T}$, $\hat{\rho}_{3}=[0,1,-R_{1}e^{-i\phi/2},0]^{T}$, and $\hat{\rho}_{4}=[0,1,-R_{2}e^{-i\phi/2},0]^{T}$ are the four basis wave functions of the left superconductor, in which $R_{1(2)}=(\epsilon\mp{i}\sqrt{\Delta^{2}-\epsilon^{2}})/\Delta$. The corresponding wave function in the right superconducting electrode is
      \begin{eqnarray}
           \psi_{R}^{S}(z)&=&D_{1}\hat{\eta}_{1}e^{ik_{S}^{+}z}+D_{2}\hat{\eta}_{2}e^{-ik_{S}^{-}z}   \label{functionSR} \\
           &&+D_{3}\hat{\eta}_{3}e^{ik_{S}^{+}z}+D_{4}\hat{\eta}_{4}e^{-ik_{S}^{-}z},   \notag
       \end{eqnarray}
      where $\hat{\eta}_{1}=[1,0,0,R_{1}e^{i\phi/2}]^{T}$, $\hat{\eta}_{2}=[1,0,0,R_{2}e^{i\phi/2}]^{T}$, $\hat{\eta}_{3}=[0,1,-R_{1}e^{i\phi/2},0]^{T}$, and $\hat{\eta}_{4}=[0,1,-R_{2}e^{i\phi/2},0]^{T}$.

      \subsection{The eigenenergy spectrum and eigenfunction of the conical ferromagnet}

      From the equation (\ref{BdG}) we obtain four eigenvalues and four eigenfunctions for our system. The first eigenfunction is determined by the expression
      \begin{equation}
          \hat{u}_{1}(z)=M_{1}\left(
          \begin{array}{c}
               e^{i(p_{1}-\frac{Q}{2})z} \\
               T_{1}e^{i(p_{1}+\frac{Q}{2})z}
          \end{array}
          \right)+M_{2}\left(
          \begin{array}{c}
               e^{i(p_{2}-\frac{Q}{2})z} \\
               T_{2}e^{i(p_{2}+\frac{Q}{2})z}
          \end{array}
          \right),       \label{function1}
      \end{equation}
       where $T_{1(2)}=-h/(\xi_{p_{1(2)}+\frac{Q}{2}}-h_{z}-\epsilon)$. The wave vectors $p_{1}$ and $p_{2}$ can be found numerically from equation $\epsilon_{1}(p_{1(2)})=\epsilon$, there the branches of the energy spectrum $\epsilon_{1(2)}(p)$ are determined by the relation (\ref{Ener_spec}).

      The second eigenfunction reads
         \begin{equation}
              \hat{u}_{2}(z)=M_{3}\left(
              \begin{array}{c}
                   T_{3}e^{i(p_{3}-\frac{Q}{2})z} \\
                   e^{i(p_{3}+\frac{Q}{2})z}
              \end{array}
              \right)+M_{4}\left(
              \begin{array}{c}
                   T_{4}e^{i(p_{4}-\frac{Q}{2})z} \\
                   e^{i(p_{4}+\frac{Q}{2})z}
              \end{array}
              \right),
         \end{equation}          \label{function2}
      where $T_{3(4)}=-h/(\xi_{p_{3(4)}-\frac{Q}{2}}+h_{z}-\epsilon)$ and the wave vectors $p_{3}$ and $p_{4}$ are the solutions of the equation $\epsilon_{2}(p_{3(4)})=\epsilon$.

      The third eigenfunction may be written as
        \begin{equation}
             \hat{v}_{1}(z)=M_{5}\left(
             \begin{array}{c}
                  e^{i(p_{5}+\frac{Q}{2})z} \\
                  T_{5}e^{i(p_{5}-\frac{Q}{2})z}
             \end{array}
             \right)+M_{6}\left(
             \begin{array}{c}
                  e^{i(p_{6}+\frac{Q}{2})z} \\
                  T_{6}e^{i(p_{6}-\frac{Q}{2})z}
             \end{array}
             \right),            \label{function3}
        \end{equation}
      where $T_{5(6)}=-h/(\xi_{p_{5(6)}-\frac{Q}{2}}-h_{z}+\epsilon)$ and the wave vectors $p_{5}$ and $p_{6}$ arise from the equation $\epsilon_{1}(p_{5(6)})=-\epsilon$.

      The fourth eigenfunction can be described as
       \begin{equation}
            \hat{v}_{2}(z)=M_{7}\left(
            \begin{array}{c}
                T_{7}e^{i(p_{7}+\frac{Q}{2})z} \\
                e^{i(p_{7}-\frac{Q}{2})z}
            \end{array}
            \right)+M_{8}\left(
            \begin{array}{c}
                T_{8}e^{i(p_{8}+\frac{Q}{2})z} \\
                e^{i(p_{8}-\frac{Q}{2})z}
            \end{array}
            \right),         \label{function4}
       \end{equation}
      where $T_{7(8)}=-h/(\xi_{p_{7(8)}+\frac{Q}{2}}+h_{z}+\epsilon)$. The corresponding wave vectors $p_{7}$ and $p_{8}$ satisfy the equation $\epsilon_{2}(p_{7(8)})=-\epsilon$. As a result, the total wave function in the ferromagnetic region can be described as
        \begin{equation}
            \psi_{F}(z)=I_{1}\otimes\hat{u}_{1}(z)+I_{1}\otimes\hat{u}_{2}(z)+I_{2}\otimes\hat{v}_{1}(z)+I_{2}\otimes\hat{v}_{2}(z),  \label{functionF}
        \end{equation}
      where $I_{1}=[1,0]^{T}$ and $I_{2}=[0,1]^{T}$.

      \subsection{Josephson current of the system}

      The wave functions [$\psi_{L}^{S}(z)$, $\psi_{F}(z)$ and $\psi_{R}^{S}(z)$] and their first derivatives should satisfy the continuity conditions at the S/F and F/S interfaces,
       \begin{equation}
           \psi_{L}^{S}(-\frac{L}{2})=\psi_{F}(-\frac{L}{2}),
           \frac{\partial\psi_{L}^{S}}{\partial{z}}\left\vert_{z=-\frac{L}{2}}\right.=\frac{\partial\psi_{F}}{\partial{z}}\left\vert_{z=-\frac{L}{2}}\right., \label{condition1}
       \end{equation}
       \begin{equation}
           \psi_{F}(\frac{L}{2})=\psi_{R}^{S}(\frac{L}{2}),
           \frac{\partial\psi_{F}}{\partial{z}}\left\vert_{z=\frac{L}{2}}\right.=\frac{\partial\psi_{R}^{S}}{\partial{z}}\left\vert_{z=\frac{L}{2}}\right.. \label{condition2}
       \end{equation}
       From these boundary conditions, we can set up 16 linear equations in the following form:
        \begin{equation}
             \hat{A}X=\hat{B},  \label{linearEq}
        \end{equation}
       where $X$ contains 16 scattering coefficients and $\hat{A}$ is a $16\times16$ matrix. The solution of the characteristic equation
      \begin{equation}
         \det\hat{A}=0    \label{characteristicEq}
      \end{equation}
       allows one to identify two Andreev bound-state solutions for energies $E_{A\sigma}$ ($\sigma $=1, 2). The Josephson current can be calculated as
       \begin{equation}
           I(\phi)=\frac{2e}{\hbar}\frac{\partial\Omega}{\partial\phi}, \label{current}
       \end{equation}
       where $\Omega$ is the phase-dependent thermodynamic potential. This potential can be obtained from the excitation spectrum by using the formula~\cite{JBardeen,JCayssol}
       \begin{equation}
           \Omega=-2T\sum_{\sigma}\ln\left[2\cosh\frac{E_{A\sigma}(\phi)}{2T}\right].   \label{potential}
       \end{equation}
       where $\Delta $, $h$, $h_{z}$, and $\mathbf{Q}$ are assumed to be the equilibrium values, which minimize the free energy of the SFS structure and depend on microscopic parameters~\cite{Buzdin-AdvPhys85}. The summation in (\ref{potential}) is taken over all positive Andreev energies [$0<E_{A\sigma}(\phi )<\Delta$]. For each value of $\phi$, we solve Eq.~(\ref{characteristicEq}) numerically to obtain the two spin-polarized Andreev levels. Since the Andreev energy spectra are doubled as they include the Bogoliubov redundancy, and only half part of the energy states should be taken into account, we can acquire the Josephson current via Eqs.~(\ref{current}) and (\ref{potential}).

      \subsection{Results and discussions}

      \begin{figure}
         \centering
         \includegraphics[width=3.45in]{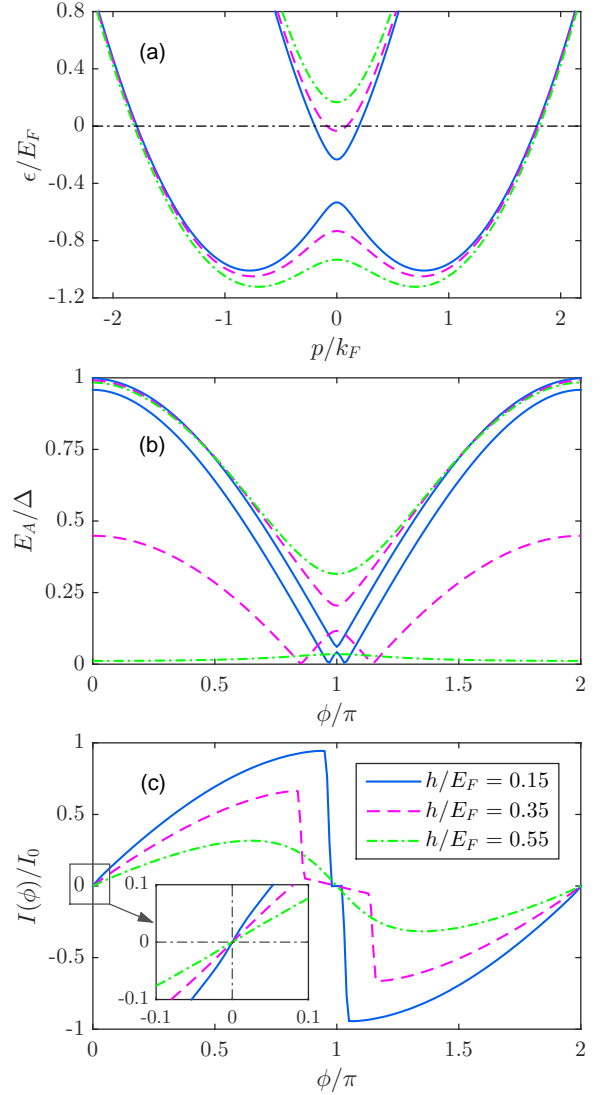} 
         \caption{(a) Energy spectrum of the helical ferromagnet, (b) Andreev bound-state energies vs the superconducting phase difference $\protect\phi$, and (c) current-phase relation for the helical ferromagnetic junction when $h/E_{F}$ takes three different values. The results plotted are for $E_{F}=1000\Delta$, $h_{z}/E_{F}=0$, $Q/k_{F}=\protect\pi/2$, and $k_{F}L=60$. The horizontal dash-dotted line in (a) denotes the Fermi level.}     \label{fig5}
      \end{figure}
       In this section, we present our results for the energy spectrum, Andreev bound-state spectrum, and the current-phase relation. Unless otherwise stated, we use the superconducting gap $\Delta$ as the unit of energy. All lengths and the exchange field strengths are measured in units of the inverse Fermi wave vector $k_{F}$ and the Fermi energy $E_{F}$, respectively. The current-phase relations are calculated at $T=0$ and the current is presented in units of $I_{0}=2e\Delta/\hbar$ {as a function of the parameters of the ferromagnetic barrier $L$, $h$, $h_{z}$, and $\mathbf{Q}$, which are supposed to be equilibrium values. Note that the different components of the exchange field produce different effects on the current-phase relations, and should be analyzed separately.

      \begin{figure}
        \centering
        \includegraphics[width=3.3in]{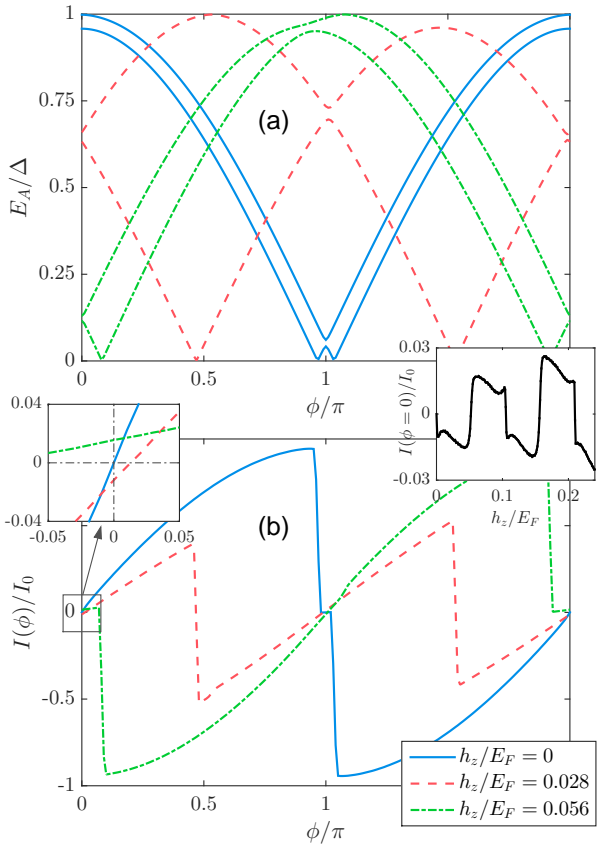} 
        \caption{(a) Andreev bound-state energies vs the superconducting phase difference $\phi$ and (b) current-phase relation for a conical ferromagnetic junction when $h_{z}/E_{F}$ takes three different values. The right inset shows the dependence of $I(\phi=0)$ on the exchange field $h_{z}/E_{F}$. The results plotted are for $E_{F}=1000\Delta$, $h/E_{F}=0.15$, $k_{F}L=60$, and $Q/k_{F}=\protect\pi/2$.} \label{fig6}
      \end{figure}

       We start our numerical solutions of the BdG equation (\ref{BdG}) from the case of the helical exchange fields $h$ without canting, i.e., for $h_{z}=0$. In Fig.~\ref{fig5}, we present the results of calculations of electrons energy spectra, Andreev bound-state spectra, and the current-phase relations for the three different values of the exchange field $h\gg\Delta$ to demonstrate the transition from the polarized metal ferromagnet to the half-metal. For chosen parameters of the F layer, the junction under consideration satisfies the short Josephson junction condition $L\ll\xi_{0}=\hbar{v_{F}}/\Delta$. For a metal interlayer, the current-phase relation is strongly nonsinusoidal and looks like the current-phase relation of short clean SNS~\cite{AAGolubov} and SFS \cite{JCayssol} junctions. In the case of the half-metal ($h/E_{F}=0.55$), the current-phase relation approaches a sinusoidal one, and as expected the critical current is strongly decreased. Note that contrary to~\cite{JCayssol}, we do not see the complete vanishing of the Josephson current in the half-metal state. As we can see in Fig.~\ref{fig5}, the Josephson current always goes to zero for $\phi=0$ and we have the standard Josephson junction behaviors in this regime.

       \begin{figure}
         \centering
         \includegraphics[width=3.4in]{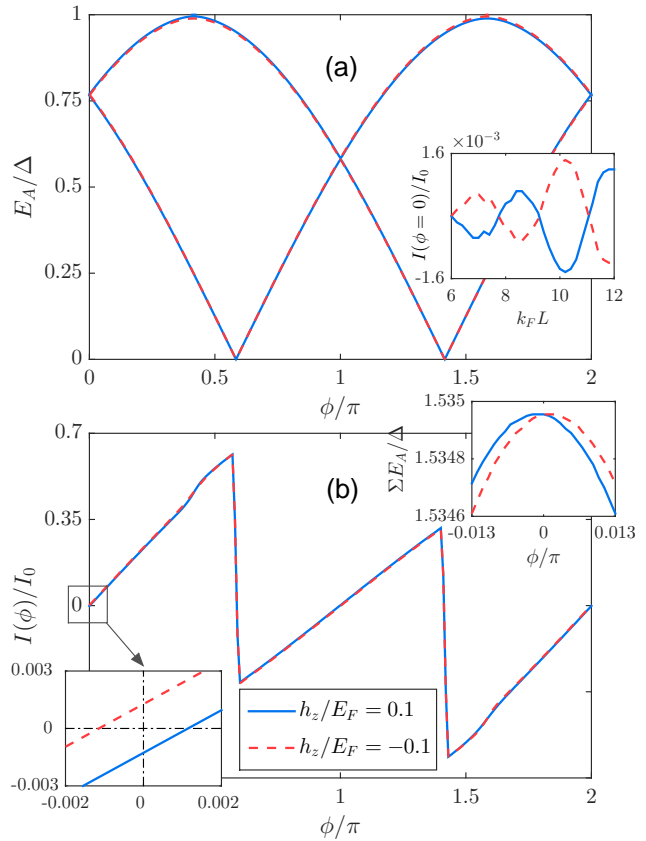} 
         \caption{(a) Andreev bound-state energies vs the superconducting phase difference $\phi$ and (b) current-phase relation for a conical ferromagnetic junction when $k_{F}L=10$. The inset in (a) shows the dependence of $I$($\phi=0$) on the thickness $k_{F}L$. The top and bottom insets in (b) illustrate the sum of the Andreev bound-state energies and the zoom of the current-phase relation near $\phi=0$, respectively. The results plotted are for $E_{F}=100\Delta $, $h/E_{F}=0.15$, and $Q/k_{F}=0.3$.}   \label{fig7}
       \end{figure}

      \begin{figure}
          \centering
          \includegraphics[width=3.3in]{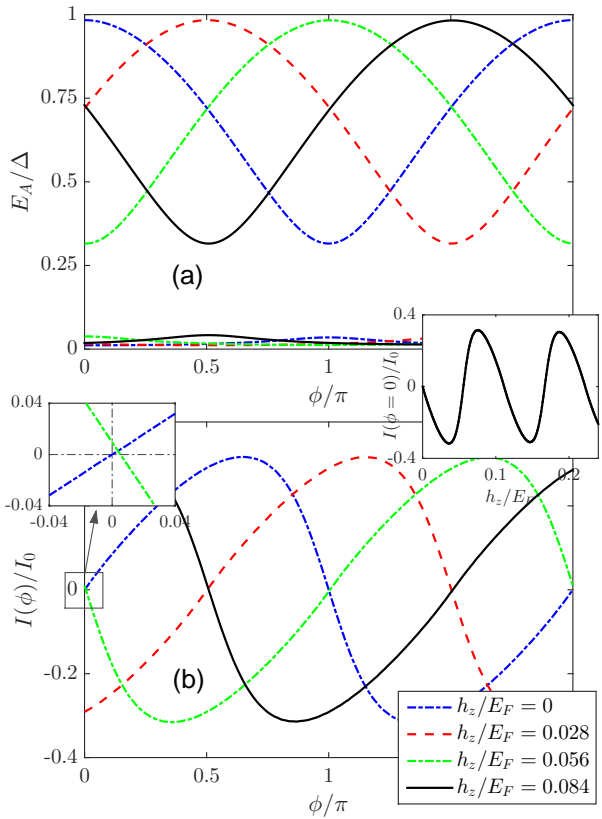} 
          \caption{(a) Andreev bound-state energies vs the superconducting phase difference $\phi$ and (b) current-phase relation for a conical half-metallic junction when $h_{z}/E_{F}$ takes several different values. The right inset shows the dependence of $I$($\protect\phi=0$) on the exchange field $h_{z}/E_{F}$. The results plotted are for $E_{F}=1000\Delta$, $h/E_{F}=0.55$, $k_{F}L=60$, and $Q/k_{F}=\pi/2$.}    \label{fig8}
      \end{figure}

      The situation changes drastically if the ferromagnetic component of the exchange field along the $z$ axis exists ($h_{z}\neq0$). Figure~\ref{fig6} shows the Andreev spectrum and the current-phase relation of a short Josephson junction with polarized ferromagnetic metal as a barrier. Small deformation of energy spectrum due to the exchange fields canting results in the qualitative modification of the Andreev spectrum and the current-phase relation: a small non zero Josephson current $I(\phi=0)$ appears in the absence of the phase difference $\phi=0$. Hence, the $\phi_{0}$ Josephson junction~\cite{IVKrive1,AAReynoso,AIBuzdin} is obtained with a finite phase difference $|\phi_{0}|\ll\pi$ in the ground state. For the exchange field $h/E_{F}<0.1$ the spontaneous current seems to be very small and the precision of our numerical analysis is not enough to study this regime. Starting at $h/E_{F}>0.1$, we clearly observe the emergence of the spontaneous current and its amplitude increase when we approach the half-metal case. The current $I(\phi =0)$ oscillates and changes sign as the canting field $h_{z}/E_{F}$ increases. For $\Delta\ll{h_{z}}\ll{h}$, the value $I(\phi=0)$ remains small in comparison with the critical current. So, the particularities of the electrons spectra in the conical ferromagnet as a weak link lead to the appearance of the spontaneous Josephson current in the absence of the phase difference. Such behavior can be understood as a phase accumulation due to the superconducting order-parameter modulation described in Sec. II.B. This modulation is proportional to $h_{z}$ in formula (\ref{max_q}) and vanishes at $h_{z}=0$.

      \begin{figure}
          \centering
          \includegraphics[width=3.3in]{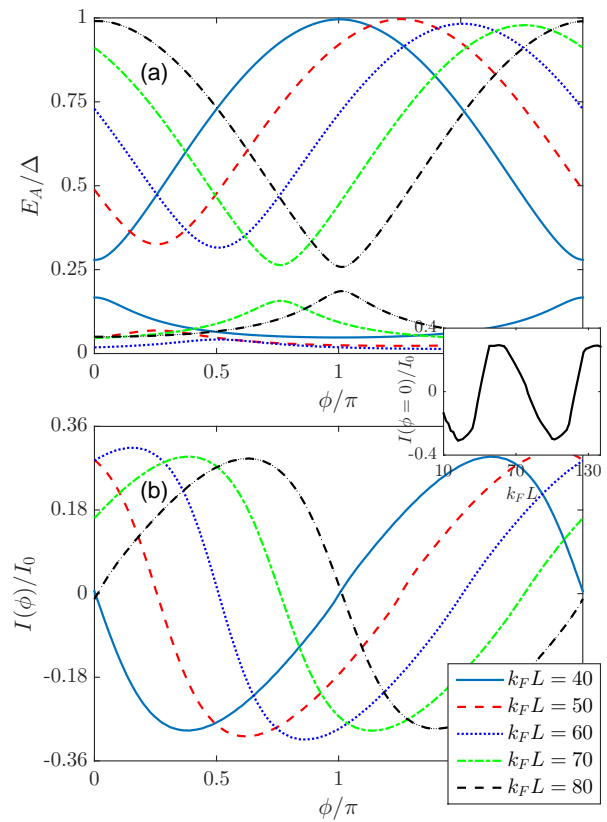} 
          \caption{(a) Andreev bound-state energies vs the superconducting phase difference $\protect\phi$ and (b) current-phase relation for a conical half-metallic junction when $k_{F}L$ varies from $40$ to $80$ with steps of $10$. The inset shows the dependence of $I$($\phi=0$) on the thickness $k_{F}L$. Here we set the parameters $E_{F}=1000\Delta $, $h/E_{F}=0.55$, $h_{z}/E_{F}=0.084$, and $Q/k_{F}=\pi/2$.} \label{fig9}
      \end{figure}

      In Fig.~\ref{fig7} we present the evolution of the Andreev spectra and the spontaneous current when the parameter $\Delta/E_{F}$ increases. The short Josephson junction condition is valid for shorter barrier $k_{F}L=10$ $(L/\xi_{0}\simeq0.05)$. A comparison of Figs.~\ref{fig6} and~\ref{fig7} shows that the Andreev spectra and the current-phase relation look similar for close values of $h_{z}/\Delta$ and $L/\xi_{0}$. The current $I(\phi=0)$ oscillates with the variation of the thickness of the ferromagnet ${L}$ and changes its sign for negative $h_{z}$: $I(\phi=0,-h_{z})=-I(\phi=0,h_{z})$ (see the inset in Fig.~\ref{fig7}). The amplitude of the spontaneous Josephson current grows as the factor $k_{F}L$ increases.

      Figures~\ref{fig8} and \ref{fig9} show how the Andreev spectrum and current-phase relation of the Josephson junction depend on the canting field $h_{z}/E_{F}$ and the barrier thickness $k_{F}L$ for the rather large ratio $h/E_{F}=0.55$, which corresponds to the half-metal state of the ferromagnet. We see that the current-phase relation for a conical half-metallic junction is close to the sinusoidal one and differs qualitatively from the previous case of the polarized ferromagnetic metal. The spontaneous current $I(\phi=0)$ and the spontaneous phase difference $\phi_{0}$ change continuously with the exchange field canting and the thickness. Hence, we can obtain a finite current at zero superconducting phase and a continuous change of the phase difference $\phi_{0}$ from $0$ to $\pi$ by tuning the exchange field canting. As expected, nonzero $h_{z}$ generates the $\phi_{0}$ junction in this case too, and the ground phase difference is very sensitive to the length of the weak link.

      \section{Conclusion}

      On the basis of the exact solution in terms of Gorkov's Green functions of the 1D model of a superconductor with a conical exchange field, we demonstrate that the ground states corresponds to the modulated superconducting phase at all temperatures. The instability of the uniform state is related to the special symmetry of the system generating the triplet superconducting correlations. We calculate the wave vector of the superconducting state modulation near the superconducting transition temperature, and we show that it is proportional to the ferromagnetic component of the conical field. These results of the exact solution are in sharp contrast to the results of the solution in the framework of the quasiclassical Eilenberger or Usadel approach, which always predict the uniform superconducting state in the case of the weak exchange field. In the second part of the article, we study the properties of the S/F/S junction with the F-conical ferromagnet. Our numerical solutions of full Bogoliubov-de Gennes equations (without the usual quasiclassical approximation) reveal the emergence of the $\phi_{0}$ junction with the finite phase difference at the ground state and nonzero current for $\phi=0$. We study how the anomalous current depends on the characteristics of the conical magnet. The revealed direct coupling between the exchange field and the Josephson phase difference paves the way for interesting implementations of the $\phi_{0}$ junctions in superconducting spintronics.

     \section*{Acknowledgments}

     The authors thank A. Melnikov and S. Mironov for useful discussions and suggestions. A.B. wishes to thank the Leverhulme Trust for supporting his stay at Cambridge University. This work was supported by French ANR project SUPERTRONICS and OPTOFLUXONICS (A.I.B.) and EU Network COST CA16218 (NANOCOHYBRI). A.V.S. acknowledges the funding from by Russian Foundation for Basic Research (Grants No. 17-52-12044 NNIO and No. 18-02-00390) and Russian Science Foundation under Grant No. 17-12-01383 (Sec. II C). H.M. acknowledges the National Natural Science Foundation of China (Grant No. 11604195) and the Youth Hundred Talents Programme of Shaanxi Province.

     \setcounter{figure}{0}
     \renewcommand{\thefigure}{A\arabic{figure}}
     \appendix

     \section{} \label{AppendA}

       The Gor'kov equations (\ref{Gorkov1}) and (\ref{Gorkov2}) can be expressed in matrix form:
       \begin{gather}
           \left(
           \begin{array}{cc}
                i\omega_{n}-\xi_{p}-h_{z} & -he^{-iQz} \\
                -he^{iQz} & i\omega_{n}-\xi_{p}+h_{z}
           \end{array}
           \right) \left(
           \begin{array}{cc}
                \hat{G}_{11} & \hat{G}_{12} \\
                \hat{G}_{21} & \hat{G}_{22}
           \end{array}
           \right)       \label{Gorkov_matrix_1} \\
           +\Delta{e^{iqz}}\left(
           \begin{array}{cc}
                0 & 1 \\
               -1 & 0
           \end{array}
           \right) \left(
           \begin{array}{cc}
                \hat{F}_{11}^{\dag} & \hat{F}_{12}^{\dag} \\
                \hat{F}_{21}^{\dag} & \hat{F}_{22}^{\dag}
           \end{array}
           \right)=\delta(\mathbf{r}-\mathbf{r}^{\prime})\left(
           \begin{array}{cc}
                1 & 0 \\
                0 & 1
           \end{array}
           \right) ,  \notag
       \end{gather}
       \begin{gather}
           \left(
           \begin{array}{cc}
                  i\omega_{n}+\xi_{p}+h_{z} & he^{iQz} \\
                  he^{-iQz} & i\omega_{n}+\xi_{p}-h_{z}
           \end{array}
           \right)\left(
           \begin{array}{cc}
                 \hat{F}_{11}^{\dag} & \hat{F}_{12}^{\dag} \\
                 \hat{F}_{21}^{\dag} & \hat{F}_{22}^{\dag}
           \end{array}
           \right)  \label{Gorkov_matrix_2}  \\
           -\Delta^{\ast}{e^{-iqz}}\left(
           \begin{array}{cc}
                0 & 1 \\
               -1 & 0
           \end{array}
           \right)\left(
           \begin{array}{cc}
                \hat{G}_{11} & \hat{G}_{12} \\
                \hat{G}_{21} & \hat{G}_{22}
           \end{array}
           \right)=0.  \notag
       \end{gather}
       Applying the Fourier transform to (\ref{Gorkov_matrix_1}) and (\ref{Gorkov_matrix_2}), we get a set of equations
       \begin{eqnarray}
          &&\left(i\omega_{n}-\xi_{p-\frac{Q}{2}+\frac{q}{2}}-h_{z}\right)\hat{G}_{11}(p-\frac{Q}{2}+\frac{q}{2},p^{\prime})  \label{Green_eq1} \\
          &&-h\hat{G}_{21}(p+\frac{Q}{2}+\frac{q}{2},p^{\prime})+\Delta\hat{F}_{21}^{\dag}(p-\frac{Q}{2}-\frac{q}{2},p^{\prime})  \notag \\
          &&=\delta(p-\frac{Q}{2}+\frac{q}{2}-p^{\prime}),   \notag
       \end{eqnarray}
       \begin{gather}
          \left(i\omega_{n}-\xi_{p+\frac{Q}{2}+\frac{q}{2}}+h_{z}\right)\hat{G}_{21}(p+\frac{Q}{2}+\frac{q}{2},p^{\prime})  \label{Green_eq2}  \\
          -h\hat{G}_{11}(p-\frac{Q}{2}+\frac{q}{2},p^{\prime})-\Delta\hat{F}_{11}^{\dag}(p+\frac{Q}{2}-\frac{q}{2},p^{\prime})=0,  \notag
       \end{gather}
       \begin{gather}
           \left(i\omega_{n}+\xi_{p+\frac{Q}{2}-\frac{q}{2}}+h_{z}\right)\hat{F}_{11}^{\dag}(p+\frac{Q}{2}-\frac{q}{2},p^{\prime}) \label{Green_eq3} \\ +h\hat{F}_{21}^{\dag}(p-\frac{Q}{2}-\frac{q}{2},p^{\prime})-\Delta^{\ast}\hat{G}_{21}(p+\frac{Q}{2}+\frac{q}{2},p^{\prime})=0,  \notag
       \end{gather}
       \begin{gather}
           \left(i\omega_{n}+\xi_{p-\frac{Q}{2}-\frac{q}{2}}-h_{z}\right)\hat{F}_{21}^{\dag}(p-\frac{Q}{2}-\frac{q}{2},p^{\prime}) \label{Green_eq4} \\
            +h\hat{F}_{11}^{\dag}(p+\frac{Q}{2}-\frac{q}{2},p^{\prime})+\Delta^{\ast}\hat{G}_{11}(p-\frac{Q}{2}+\frac{q}{2},p^{\prime})=0.  \notag
       \end{gather}
       The solutions of (\ref{Green_eq1})-(\ref{Green_eq4}) provide the expression for $\hat{F}_{11}^{\dag}(p+\frac{Q}{2}-\frac{q}{2},p^{\prime})$, $\hat{F}_{21}^{\dag}(p-\frac{Q}{2}-\frac{q}{2},p^{\prime})$, $\hat{G}_{11}(p-\frac{Q}{2}+\frac{q}{2},p^{\prime})$ and $\hat{G}_{21}(p+\frac{Q}{2}+\frac{q}{2},p^{\prime})$. Following the same derivation procedure, we can get another set of equations from (\ref{Gorkov_matrix_1}) and (\ref{Gorkov_matrix_2}) for Green functions $\hat{F}_{22}^{\dag}(p-\frac{Q}{2}-\frac{q}{2},p^{\prime})$, $\hat{F}_{12}^{\dag}(p+\frac{Q}{2}-\frac{q}{2},p^{\prime})$, $\hat{G}_{22}(p+\frac{Q}{2}+\frac{q}{2},p^{\prime})$ and $\hat{G}_{12}(p-\frac{Q}{2}+\frac{q}{2},p^{\prime})$. These equations coincide with (\ref{Green_eq1})-(\ref{Green_eq4}) provided ($\hat{F}_{11}$, $\hat{F}_{21}$, $\hat{G}_{11}$, $\hat{G}_{21}$) are replaced by ($-\hat{F}_{22}$, $-\hat{F}_{12}$, $\hat{G}_{22}$, $\hat{G}_{12}$) and ($\omega_{n}$, $Q$, $q$, $h$, $h_{z}$) are replaced by ($\omega_{n}$, $-Q$, $q$, $h$, $-h_{z}$).

     \section{} \label{AppendB}

       To obtain $\hat{F}_{21}^{\dag}$ in a linear-over-$\Delta$ approximation, it is sufficient to neglect the quadratic term $\left\vert\Delta\right\vert^{2}$ in Eqs.~(\ref{F21}) and (\ref{DW}). Performing the expansion over $h^{2}$ and also making the substitutions $Q/2\rightarrow\tilde{Q}$ and $q/2\rightarrow\tilde{q}$, the expressions of $\hat{F}_{21}^{\dag}$ can be simplified into the following form:
       \begin{eqnarray}
           &&\hat{F}_{21}^{\dag}\left(p-\tilde{Q}-\tilde{q},p^{\prime}\right)=-\delta\left(p-\tilde{Q}+\tilde{q}-p^{\prime}\right)\tilde{F}_{21}^{\dag}\,,\label{F21Ap1}\\
           &&\tilde{F}_{21}^{\dag }=-\frac{(A_{3}A_{4}+h^{2})\Delta^{\ast}}{A_{1}A_{2}A_{3}A_{4}-(A_{1}A_{3}+A_{2}A_{4})h^{2}} \label{F21_App} \\
           &&\qquad\simeq-\frac{\Delta^{\ast}}{A_{1}A_{2}}{(1+\frac{h^{2}}{A_{3}A_{4}}+\frac{h^{2}}{A_{2}A_{4}}+\frac{h^{2}}{A_{1}A_{3}})}\,,  \notag
       \end{eqnarray}
       where $A_{1}\sim{A_{4}}$ are determined by the expressions
       \begin{eqnarray*}
            A_{1}&=&i\omega_{n}-\xi\left(p-\tilde{Q}+\tilde{q}\right)-h_{z}\simeq{i}\omega_{n}-\xi+X_{1}, \\
            A_{2}&=&i\omega_{n}-\xi\left(p+\tilde{Q}+\tilde{q}\right)+h_{z}\simeq{i}\omega_{n}+\xi+X_{2}, \\
            A_{3}&=&i\omega_{n}+\xi\left(p+\tilde{Q}-\tilde{q}\right)+h_{z}\simeq{i}\omega_{n}-\xi+X_{3}, \\
            A_{4}&=&i\omega_{n}+\xi\left(p-\tilde{Q}-\tilde{q}\right)-h_{z}\simeq{i}\omega_{n}+\xi+X_{4}.
       \end{eqnarray*}
       Here $\xi=p^{2}/2m-E_{F}$ and
       \begin{eqnarray*}
            &&X_{1}=v(\tilde{Q}-\tilde{q})+\frac{\tilde{Q}\tilde{q}}{m}-h_{z}\,,X_{2}=-v(\tilde{Q}+\tilde{q})+\frac{\tilde{Q}\tilde{q}}{m}-h_{z}\,,\\
            &&X_{3}=-v(\tilde{Q}+\tilde{q})-\frac{\tilde{Q}\tilde{q}}{m}+h_{z}\,,X_{4}=v(\tilde{Q}-\tilde{q})-\frac{\tilde{Q}\tilde{q}}{m}+h_{z}\,.
       \end{eqnarray*}

       As a result, the function $\tilde{F}_{21}^{\dag}$ can be expressed as
       \begin{eqnarray}
            \tilde{F}_{21}^{\dag}&=&-\Delta^{\ast}\left[\frac{1}{(i\omega_{n}-\xi+X_{1})(i\omega_{n}+\xi+X_{2})}\right. \label{F21_induce} \\
            &&+\frac{1}{(i\omega_{n}-\xi+X_{1})(i\omega_{n}+\xi+X_{2})}  \notag \\
            &&\times\frac{1}{(i\omega_{n}-\xi +X_{3})(i\omega_{n}+\xi+X_{4})}  \notag \\
            &&+\frac{1}{(i\omega_{n}-\xi+X_{1})(i\omega_{n}+\xi+X_{2})^{2}(i\omega_{n}+\xi+X_{4})}  \notag \\
            &&\left.+\frac{1}{(i\omega_{n}-\xi+X_{1})^{2}(i\omega_{n}+\xi+X_{2})(i\omega_{n}-\xi+X_{3})}\right] \,.  \notag
       \end{eqnarray}
       Performing the integration over $\xi$ in (\ref{F21_induce}), we find
       \begin{eqnarray}
           &&\int\frac{\tilde{F}_{21}^{\dag}}{\Delta^{\ast}}d\xi\simeq2\pi{i}\left\{\frac{1}{2\left(i\omega_{n} -v\tilde{q}+\frac{\tilde{Q}\tilde{q}}{m}-h_{z}\right)}\right.   \label{InF21Ap} \\
           &&+\frac{h^{2}(i\omega_{n}-v\tilde{q})}{4\left[\left(i\omega_{n}-v\tilde{q}\right)^{2}-\left(\frac{\tilde{Q}\tilde{q}}{m} \notag
           -h_{z}\right)^{2}\right]\left[\left(i\omega_{n}-v\tilde{q}\right)^{2}-v^{2}\tilde{Q}^{2}\right]}\\ \notag
           &&+\frac{h^{2}}{8\left(i\omega_{n}-v\tilde{q}+\frac{\tilde{Q}\tilde{q}}{m}-h_{z}\right)^{2}\left(i\omega_{n} +v\tilde{Q}-v\tilde{q}\right)}\\\notag
           &&+\left.\frac{h^{2}}{8\left(i\omega_{n}-v\tilde{q}+\frac{\tilde{Q}\tilde{q}}{m} -h_{z}\right)^{2}\left(i\omega_{n}-v\tilde{Q}-v\tilde{q}\right)}\right\}.  \notag
       \end{eqnarray}
       If one performs the Taylor expansion of (\ref{InF21Ap}) to the second power of $\tilde{q}$ in the limit $h\ll{T_{c0}}$, the equation for the critical temperature becomes
       \begin{eqnarray}
            &&\ln\left(\frac{T_{c}}{T_{c0}}\right)=2\pi{T_{c}}\sum_{\omega_{n}\geq0}\left\{\frac{\omega_{n}}{\omega_{n}^{2} +h_{z}^{2}}-\frac{1}{\omega_{n}}\right. \label{Tcx}  \\
            &&-\frac{\omega_{n}^{3}h^{2}}{\left(\omega_{n}^{2}+h_{z}^{2}\right)^{2}\left(\omega_{n}^{2} +v^{2}\tilde{Q}^{2}\right)}-\frac{4\omega_{n}^{3}\tilde{Q}h^{2}h_{z}\tilde{q}}{m\left(\omega_{n}^{2} +h_{z}^{2}\right)^{3}\left(\omega_{n}^{2}+v^{2}\tilde{Q}^{2}\right)}  \notag \\
            &&\left.+\frac{\omega_{n}v^{2}\left(3h_{z}^{2}-\omega_{n}^{2}\right)\tilde{q}^{2}}{\left(\omega_{n}^{2}+h_{z}^{2}\right)^{3}}\right\}.\notag
       \end{eqnarray}

       Using the definition $\Delta{T_{c}}=T_{c}-T_{c0}$ and the relation $\ln\left(\frac{T_{c}}{T_{c0}}\right)\approx\frac{\Delta{T_{c}}}{T_{c0}}$, in the limit $h_{z}\ll{T_{c0}}$ we have
       \begin{eqnarray}
            \frac{\Delta{T_{c}}}{T_{c0}}&=&2\pi{T_{c}}\sum_{\omega_{n}\geq0}\left[-\frac{h_{z}^{2}}{\omega_{n}^{3}} -\frac{h^{2}}{\left(\omega_{n}^{2}+v^{2}\tilde{Q}^{2}\right)\omega_{n}}\right. \\
            &&\left.-\frac{4\tilde{Q}h^{2}h_{z}\tilde{q}}{m\left(\omega_{n}^{2}+v^{2}
            \tilde{Q}^{2}\right)\omega_{n}^{3}}-\frac{v^{2}\tilde{q}^{2}}{\omega_{n}^{3}}\right].  \notag
       \end{eqnarray}
       Finally, by the opposite substitutions ${\tilde{Q}}\rightarrow\frac{Q}{2}$ and ${\tilde{q}}\rightarrow\frac{q}{2}$ we obtain
       \begin{eqnarray}
            \frac{\Delta{T_{c}}}{T_{c0}}&=&2\pi{T_{c}}\sum_{\omega_{n}\geq0}\left[-\frac{h_{z}^{2}}{\omega_{n}^{3}} -\frac{4h^{2}}{\left(4\omega_{n}^{2}+v^{2}Q^{2}\right)\omega_{n}}\right. \\
            &&\left.-\frac{4Qh^{2}h_{z}q}{m\left(4\omega_{n}^{2}+v^{2}Q^{2}\right)\omega_{n}^{3}}-\frac{v^{2}q^{2}}{4\omega_{n}^{3}}\right].  \notag
       \end{eqnarray}

     \section{} \label{AppendC}

       From (\ref{Green_eq1})--(\ref{Green_eq4}) we get the Green's function $\hat{G}_{11}(p,p^{\prime})$ for the uniform superconductor ($q$=0) with a helical magnetic order
       \begin{eqnarray}
           &&\hat{G}_{11}(p,p^{\prime})=\delta(p-p^{\prime})    \label{G11-AppC} \\
           &&\times\left[\frac{(i\omega-\xi_{p+Q}+h_{z})(i\omega+\xi_{p+Q}+h_{z})(i\omega+\xi_{p}-h_{z})}{D_{1}(\omega)}\right.   \notag \\
           &&\left.-\frac{(i\omega-\xi_{p+Q}+h_{z})h^{2}+(i\omega+\xi_{p}-h_{z})\left\vert\Delta\right\vert^{2}}{D_{1}(\omega)}\right],\notag
       \end{eqnarray}
       where $D_{1}(\omega)=$
       \begin{eqnarray}
            &&\left[(i\omega-\xi_{p}-h_{z})(i\omega+\xi_{p}-h_{z})+h^{2}-\left\vert\Delta\right\vert^{2}\right]  \label{D1W} \\
            &&\times\left[(i\omega-\xi_{p+Q}+h_{z})(i\omega+\xi_{p+Q}+h_{z})+h^{2}-\left\vert\Delta\right\vert^{2}\right]\notag\\
            &&-(2i\omega-\xi_{p}+\xi_{p+Q})(2i\omega-\xi_{p+Q}+\xi_{p})h^{2},  \notag
       \end{eqnarray}
       $\xi_{p}=\xi(p)=p^{2}/2m-E_{F}$, and we use $\omega$ instead of $\omega_{n}$ for short. The solutions for the Green function $\hat{G}_{22}$ are described by the same expressions (\ref{G11-AppC}) and (\ref{D1W}) by replacing $Q\rightarrow-Q$ and $h_{z}\rightarrow-h_{z}$. Taking into account the symmetry relation between the Green functions
       \begin{equation*}
           \hat{G}_{11}(-p,-h_{z})=\hat{G}_{22}(p,h_{z})\,,
       \end{equation*}
       the supercurrent in a magnetic superconductor with spiral magnetic order,
       \begin{equation}
           J=\left.\frac{ie}{m}(\nabla_{r^{\prime}}-\nabla_{r})\left[\hat{G}_{11}(r,r^{\prime})+\hat{G}_{22}(r,r^{\prime})\right]\right\vert_{r^{\prime}\rightarrow{r}}, \label{CurDen-Gr}
       \end{equation}
       can be written via the Green function $\hat{G}_{11}$ (\ref{G11-AppC}) as follows:
       \begin{equation}
           J=\frac{2e}{m}\iint\,dp\,d\omega\left[p\,\hat{G}_{11}(p,h_{z})-p\,\hat{G}_{11}(p,-h_{z})\right]. \label{CurDen-G11}
       \end{equation}
       Although it is possible to carry out these calculations for arbitrary $h_{z}$, we restrict our consideration to only terms linear on $h_{z}$ in $\hat{G}_{11}$. In this case the expression (\ref{G11-AppC}) can be expanded into the following form:
       \begin{equation}
          \hat{G}_{11}(p,p^{\prime})=\hat{G}_{11}^{(0)}(p,p^{\prime})+h_{z}\hat{G}_{11}^{(1)}(p,p^{\prime})+\Delta\hat{G}_{11}(p,p^{\prime}) \label{G11-hz}
       \end{equation}
       where
       \begin{equation}
           \hat{G}_{11}^{(0)}(p,p^{\prime})=\delta(p-p^{\prime})\frac{-\xi_{p}(\omega^{2} +\xi_{p+Q}^{2}+\left\vert\Delta\right\vert^{2})-\xi_{p+Q}h^{2}}{\left[\,\omega^{2}+E_{1}^{2}\,\right]\left[\,\omega^{2}+E_{2}^{2}\,\right]} \label{G11-0}
       \end{equation}
       and
       \begin{eqnarray}
            &&\hat{G}_{11}^{(1)}(p,p^{\prime})=\delta(p-p^{\prime})\left[\frac{\xi_{p+Q}^{2}-\omega^{2}-h^{2}+\left\vert\Delta\right\vert^{2}}{\left[ \,\omega^{2}+E_{1}^{2}\,\right]\left[\,\omega^{2}+E_{2}^{2}\,\right]}\right.   \label{G11-1} \\
            &&\left.-2\omega^{2}\frac{(\xi_{p+Q}^{2}-\xi_{p}^{2})(\omega^{2}+\xi_{p+Q}^{2}+\left\vert\Delta\right\vert^{2}+h^{2})}{\left[ \,\omega^{2}+E_{1}^{2}\,\right]^{2}\left[\,\omega^{2}+E_{2}^{2}\right]^{2}}\,\right]\,.  \notag
       \end{eqnarray}
       The last item $\Delta\hat{G}_{11}(p,p^{\prime})$ in (\ref{G11-hz}) includes terms that are odd in frequency $\omega$, which does not contribute to the integral $\int\,d\omega ...$, and/or terms containing a higher power of $h_{z}$. The significant components $\hat{G}_{11}^{(0)}(p,p^{\prime})$ and $\hat{G}_{11}^{(1)}(p,p^{\prime})$ are described by the energy spectra
       \begin{eqnarray}
            E_{1,2}^{2}&=&\zeta^{2}+\eta^{2}+\left\vert\Delta\right\vert^{2}+h^{2}\\ \label{E1E2}
            &&\pm2\sqrt{\zeta^{2}\left(\eta^{2}+h^{2}\right)+\left\vert\Delta\right\vert^{2}h^{2}}\,,  \notag
       \end{eqnarray}
       where $\zeta=\left(\xi_{p}+\xi_{p+Q}\right)/2$ and $\eta=\left(\xi_{p}-\xi_{p+Q}\right)/2$.

       \begin{figure}
          \centering
          \includegraphics[width=3.3in]{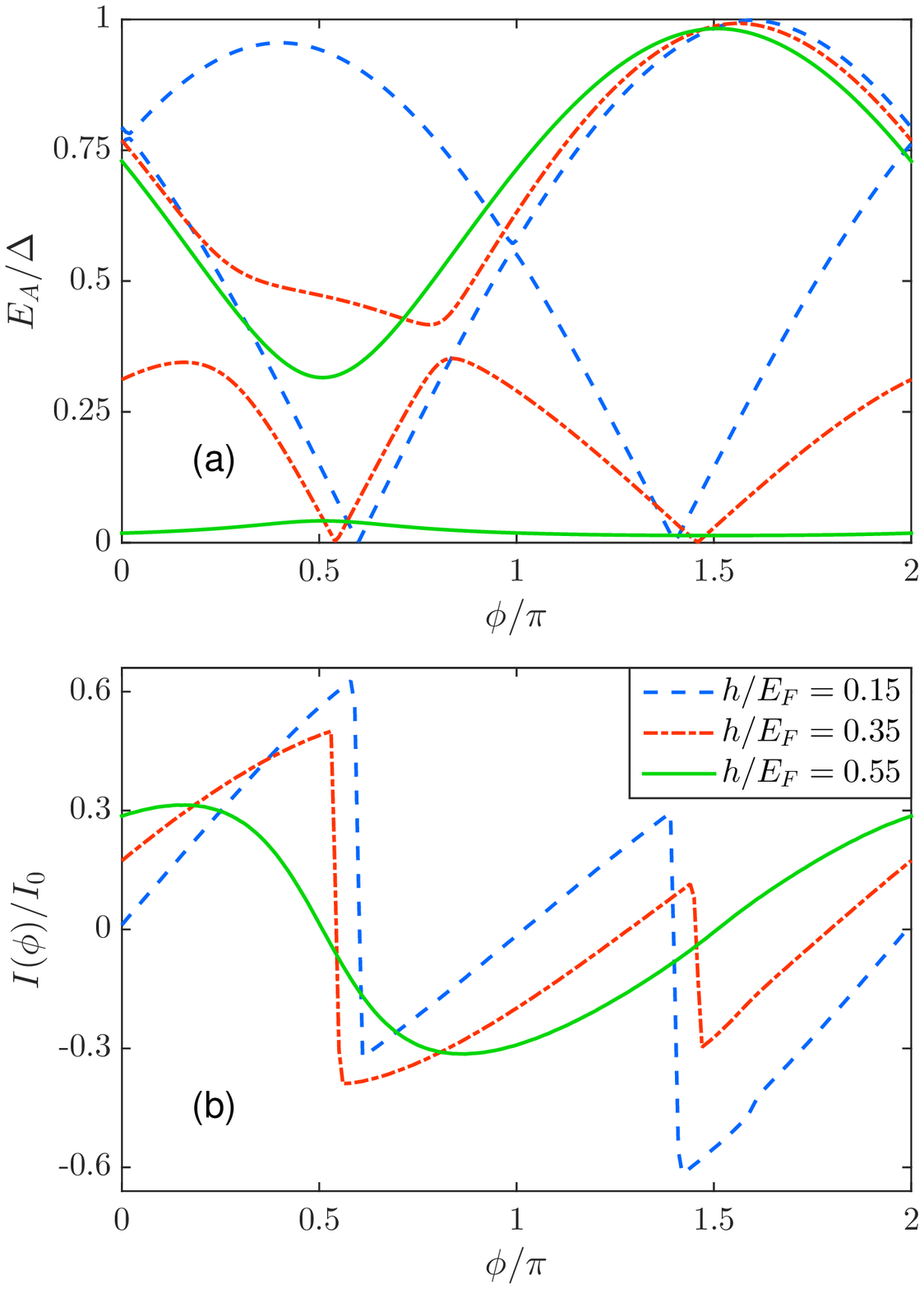} 
          \caption{(a) Andreev bound-state energies vs the superconducting phase difference $\phi$ and (b) current-phase relation for the conical ferromagnetic junction when $h/E_{F}$ takes three different values. The results plotted are for $k_{F}L=60$, $E_{F}=1000\Delta$, $h_{z}/E_{F}=0.084$, and $Q/k_{F}=\protect\pi/2$.}   \label{fig_a1}
       \end{figure}
       
       \begin{figure}
          \centering
          \includegraphics[width=3.3in]{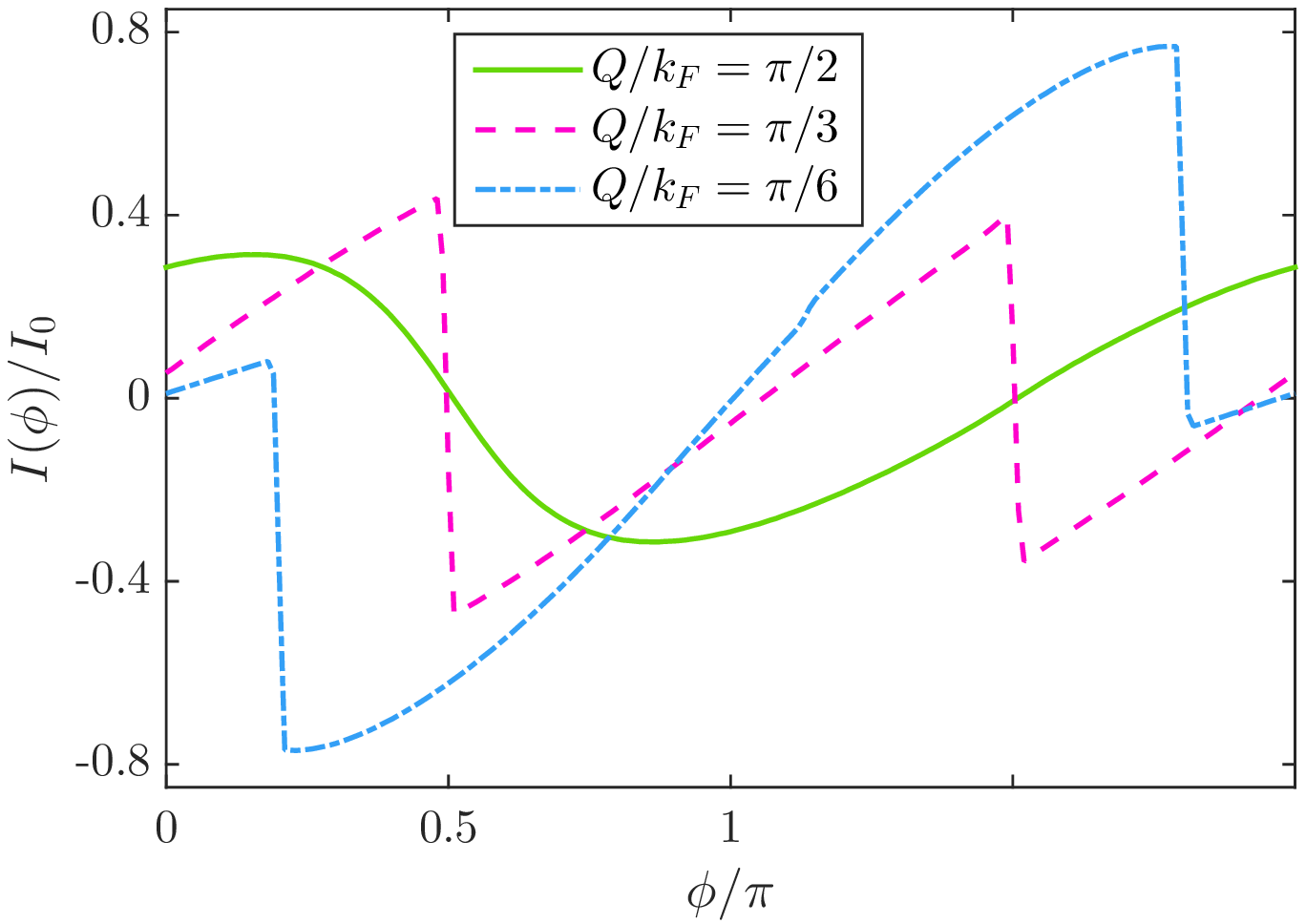} 
          \caption{Current-phase relation for the conical ferromagnetic junction when $Q$ takes three different values. We set the parameters $E_{F}=1000\Delta$, $h/E_{F}=0.55$, $h_{z}/E_{F}=0.084$, and $k_{F}L=60$.} \label{fig_a2}
       \end{figure}

       Substituting expansion (\ref{G11-hz}) into Eq.~(\ref{CurDen-G11}), we get
       \begin{equation}
            J=\frac{4eh_{z}}{m}\iint\,dp\,d\omega\left[\left(p-Q/2\right)\hat{G}_{11}^{(1)}\left(p-Q/2,p^{\prime}\right)\right]. \label{CurDen-G11-1}
       \end{equation}
       Performing long but straightforward calculations, we find the following analytical expression for supercurrent (\ref{CurDen-G11-1}),
       \begin{eqnarray}
            J&=&\frac{8e\pi\tilde{Q}\tilde{h}_{z}}{m}\left\{\int{d}\tilde{p}\frac{\tilde{p}^{2} \tilde{\xi}_{Q}(\tilde{p})}{e_{1}^{2}e_{2}+e_{1}e_{2}^{2}}\right.  \label{CurDen-ApC}  \\
            &+&\frac{1}{4}\int\,d\tilde{p}\,\frac{2\tilde{h}^{2}-\left(e_{1}-e_{2}\right)^{2}/2}{e_{1}^{2}e_{2}+e_{1}e_{2}^{2}}  \notag \\
            &+&\left.\int{d}\tilde{p}\,\frac{\tilde{p}^{2}\tilde{\xi}_{Q}(\tilde{p})\left[2\tilde{Q}^{2}\tilde{\xi}_{Q}(\tilde{p}) -(e_{1}^{2}+e_{2}^{2})\right]}{e_{1}e_{2}(e_{1}+e_{2})^{3}}\right\},  \notag
       \end{eqnarray}
        where $e_{1,2}=E_{1,2}/E_{F}$ and $\tilde{\xi}_{Q}(\tilde{p})=\tilde{p}^{2}+\tilde{Q}^{2}/4-1$. Here we use the dimensionless variables $\tilde{h}$, $\tilde{h}_{z}$, and $\tilde{\Delta}$ in the units of Fermi energy $E_{F}=p_{F}^{2}/2m$ as well as $\tilde{Q}$, $\tilde{p}$ in the units of Fermi momentum $p_{F}$.

      \section{} \label{AppendD}

       In Fig.~\ref{fig_a1} we plot the Andreev spectrum and the current-phase relation for increasing exchange fields $h/E_{F}$ when the energy band structure changes from ferromagnet to half-metal. We note that with an increase of $h/E_{F}$, the asymmetry of the Andreev spectrum structure is enhanced and the phase shift $\phi_{0}$ increases accordingly.

       In Fig.~\ref{fig_a2} it is shown how the transition from ferromagnet to half-metal with the increase of the helical modulation vector $Q$ changes the spontaneous current.

    \end{document}